\documentclass[preprint,12pt]{elsarticle}

\usepackage[dvipsnames]{xcolor}
\usepackage[colorlinks,linkcolor=blue,urlcolor=blue,citecolor=blue]{hyperref}
\usepackage{scalefnt}

\newcommand{\changed}[1]{#1}

\usepackage{listings}
\usepackage{tikz}
\usetikzlibrary{shapes,arrows,positioning,calc}
\usepackage{pgfplotstable}

\usepackage{graphicx}

\usepackage{amsmath}
\usepackage{amsfonts}
\usepackage{amssymb}
\usepackage{algorithm}

\usepackage{floatflt}
\usepackage{siunitx}

\usepackage{pdfsync}

\usepackage{lineno}

\usepackage{todonotes}

\newcommand{\R}{{\mathbb R}}

\newcommand{\class}[1]{\texttt{#1}}
\newcommand{\library}[1]{\texttt{#1}}
\newcommand{\application}[1]{\texttt{#1}}

\newcommand{\dealii}{{\textsc{deal.II}}}

\newcommand{\trilinos}{{\textsc{Trilinos}}}

\usepackage[draft,inline,marginclue,multiuser]{fixme}

\FXRegisterAuthor{da}{ada}{DA} 
\FXRegisterAuthor{wb}{aom}{WB} 
\FXRegisterAuthor{dd}{asg}{DD}
\FXRegisterAuthor{th}{ath}{TH}
\FXRegisterAuthor{lh}{alh}{LH}
\FXRegisterAuthor{mk}{amk}{MK}
\FXRegisterAuthor{mm}{amm}{MM}
\FXRegisterAuthor{jp}{ajp}{JP}
\FXRegisterAuthor{bt}{abt}{BT}
\FXRegisterAuthor{dw}{adw}{DW}

\tikzset{
  block/.style = {fill=white, minimum height=3em, minimum width=3em},
  core/.style={
           rectangle,
           rounded corners,
           draw=black, thick,
           minimum height=1.5em,
           inner sep=2pt,
           text centered,
         },
   optional/.style={
           rectangle,
           rounded corners,
           draw=black, thick,
           minimum height=1.5em,
           inner sep=2pt,
           text centered,
         },
         library/.style={
           rectangle,
           draw=black, thin, dashed,
           minimum height=1.5em,
           inner sep=2pt,
           text centered,
         },
         application/.style={
           rectangle,
           rounded corners,
           draw=black, thin, dashed,
           minimum height=1.5em,
           inner sep=2pt,
           text centered,
         },
    line-optional/.style={
    draw=black, thin, dashed
    },
    line-important/.style={
    draw=black, thick
    },
}

\biboptions{compress}

\journal{Computers \& Mathematics with Applications}

\begin{document}


\begin{frontmatter}

\title{\changed{The \textsc{deal.II} finite element library: \\Design,
  features, and insights}}

\author{Daniel Arndt
\footnotetext{
  This manuscript has been authored by UT-Battelle, LLC under Contract No.
  DE-AC05-00OR22725 with the U.S. Department of Energy. The United States
  Government retains and the publisher, by accepting the article for
  publication, acknowledges that the United States Government retains a
  non-exclusive, paid-up, irrevocable, worldwide license to publish or
  reproduce the published form of this manuscript, or allow others to do
  so, for United States Government purposes. The Department of Energy will
  provide public access to these results of federally sponsored research in
  accordance with the DOE Public Access Plan
  (http://energy.gov/downloads/doe-public-access-plan).}}
\address{Computational Engineering and Energy Sciences Group,
         Oak Ridge National Laboratory;
         Oak Ridge, TN, USA}

\author{Wolfgang Bangerth}
\address{Department of Mathematics,
  Colorado State University;
  Fort Collins, CO, USA}

\author{Denis Davydov}
\address{Independent Researcher}

\author{Timo Heister}
\address{School of Mathematical and Statistical Sciences,
  Clemson University;
  Clemson, SC, USA}

\author{Luca Heltai}
\address{International School for Advanced Studies, Trieste, TS, Italy}

\author{Martin Kronbichler}
\address{Institute for Computational Mechanics,
  Technical University of Munich;
  Garching, Germany}

\author{Matthias Maier}
\address{Department of Mathematics,
  Texas A\&M University;
  College Station, TX, USA}

\author{Jean-Paul Pelteret}
\address{Independent Researcher}

\author{Bruno Turcksin}
\address{Computational Engineering and Energy Sciences Group,
  Oak Ridge National Laboratory;
  Oak Ridge, TN, USA}

\author{David Wells}
\address{Department of Mathematics,
  University of North Carolina, Chapel Hill;
  Chapel Hill, NC, USA}

\begin{abstract}
    \dealii{} is a state-of-the-art finite element library focused on
    generality, dimension-independent programming, parallelism, and
    extensibility. Herein, we outline its primary design considerations and
    its sophisticated features such as distributed meshes, $hp$-adaptivity,
    support for complex geometries, and matrix-free algorithms. But
    \dealii{} is more than just a software library: It is also a diverse
    and worldwide community of developers and users, as well as an
    educational platform. We therefore also discuss some of the technical
    and social challenges and lessons learned in running a large community
    software project over the course of two decades.
\end{abstract}


\end{frontmatter}


\section{Introduction}
\label{sec:introduction}

Mathematical software has been collected in packages for almost as long as
computers have been around. The first of these packages were collections of
loosely connected subroutines for specific purposes. In the earliest days,
most of these were related to linear algebra problems such as the solution
of linear systems, or computing eigenvalues, but also to numerical
integration and differentiation. Few of these packages survive to this day,
but the BLAS and related LAPACK interfaces
\cite{Lawson:1979:BLA:355841.355847,laug} are still widely used, despite
the fact that BLAS was developed and standardized already in the 1970s.

Since then, mathematical software has seen the emergence of ever more
sophisticated and connected libraries. This includes software for sparse
linear algebra in the 1980s; support for parallel sparse linear algebra
based on MPI \cite{MPI15} in the 1990s; and, since the late 1990s and early
2000s, libraries that provide the tools to build numerical solvers for
partial differential equations (PDEs) using finite element, finite
difference, or finite volume methods. Many of the libraries in this
category are discussed in articles in this issue.

Among the largest of the libraries supporting numerical PDE solvers is
\dealii{}, whose architecture, feature set, user and developer community,
and applications we discuss herein. The origins of this library lie in the
Numerical Analysis Group at the University of Heidelberg, Germany, where a
predecessor library called DEAL (short for the ``Differential Equation
Analysis Library'') was developed since the mid-1990s. \dealii{} is a
re-write of DEAL using more modern software design principles; it was
started in late 1997 by Wolfgang Bangerth, Ralf Hartmann, and Guido
Kanschat who, at the time,  were all members of the same group in
Heidelberg. Since then, \dealii{} has grown into a truly worldwide project
with more than one million lines of C++ code, to which more than 250 people
have contributed, and that is managed by a dedicated group of Principal
Developers located at universities, research institutes, and companies
across continents.

This paper discusses aspects of the \dealii{} project. Specifically,
Section~\ref{sec:design} is concerned with design considerations that
dictate the functionality that \dealii{} provides.
Section~\ref{sec:features} then covers specific functionality provided
within this framework. As will become apparent there, our goal is to cover
essentially everything that can be provided in a generic way to codes that
want to solve specific partial differential equations using the most modern
aspects of the finite element method, all while supporting modern hardware.
Section~\ref{sec:lessons} discusses some of the lessons we have learned
running a large and complex software project, while
Section~\ref{sec:education} covers how \dealii{} supports our views on, and
activities in education in the Computational Science and Engineering arena.
Section~\ref{sec:applications} briefly outlines some of the complex
applications that have been built atop \dealii{} over the years
\changed{
 and 
showcases new parallel scalability results. In 
Section~\ref{sec:future} we comment on the vision and future directions
of the project.
}
We
conclude in Section~\ref{sec:conclusions}.

We end this introduction by stating that an earlier review of \dealii{} was
previously published in \cite{BangerthHartmannKanschat2007}, and that
individual releases and new features are discussed in a series of papers of
which the most recent ones are \cite{dealII84,dealii85,dealii90,dealII91}.
Specific features of \dealii{}, along with details of their implementation,
are discussed in a large number of papers
\cite{Kanschat2004,
  JanssenKanschat2011,
  clevenger_par_gmg,
  BangerthBursteddeHeisterKronbichler11,
  BangerthKayserHerold2007,
  Davydov2016,
  KronbichlerKormann2012,
  KronbichlerKormann2019,
  DeSimoneHeltaiManigrasso2009,
  GiulianiMolaHeltai-2018-a,
  MaierBardelloniHeltai-2016-a,MaierBardelloniHeltai-2016-b,
  TKB16,
  SartoriGiulianiBardelloni-2018-a}.
For more details and an updated list, see
\url{https://dealii.org/publications.html} or the summary in
\cite{dealII91}.


\section{Design considerations}
\label{sec:design}

Any nontrivial software package needs (written or unwritten) design
principles to guide its development. Such principles provide a mental
backdrop for expectations on how its components are used and interact with
each other. Design principles also enable users to learn a software package
efficiently, build a foundation for the evolution of the software, and aid
developers in gauging an appropriate and idiomatic implementation of new
features.

In the following subsections, we outline the design principles upon which
\dealii{} is built today. Some of these principles were already present at
the start in 1997 -- as explicit design goals of what we wanted to achieve
at the time --, whereas others developed over time: implicitly at first,
and explicitly codified as part of our development practices later on.

\subsection{A complete toolbox for finite element codes}

Finite element codes are often large and complex. They use many pieces of
functionality, including meshes, geometry descriptions, shape functions,
mappings, quadratures, linear algebra representations and algorithms, and
more. As a consequence, such codes can run into tens or hundreds of
thousands of lines of code when written from scratch.

\dealii{} strives to provide all functionality related to the finite
element discretization of partial differential equations: an extensive
collection of tools that are \emph{generic} with respect to the
discretization of any one partial differential equation, while
\emph{leaving the decision about how to put these pieces together} to the
user who can combine them freely in their application codes.

\subsection{No hidden magic: \dealii{} is a library, not a framework}
\label{sec:no-magic}

Computational software packages can roughly be categorized as either
libraries or frameworks. A library is a collection of building blocks (data
structures and algorithms that work on them) that can be combined in more
or less arbitrary ways in a user program that \textit{builds on} the
library and that typically provides the overall logic and outer loops. One
can think of MATLAB as an examplar of a library in this sense; BLAS and
LAPACK are more traditional examples. On the other hand, frameworks provide
the overall logic and let users fill in specific pieces. Many solid
mechanics software packages are of this kind: they implement the overall
solution algorithms and users only have to describe specifics such as the
geometry, boundary conditions, loading forces, and the details of the
material constitutive laws. Frameworks are therefore often easy to use but
are restricted to specific purposes: It is easy to replace one material
description by another, but it might be impossible to implement a
dual-weighted error estimator requiring the solution of an adjoint problem
since this would require changing the outermost logic, which may not be
accessible to users.

\dealii{} is a library in this dichotomy. It strives to provide all tools
our users may need to write efficient and flexible finite element programs,
but it does not dictate the overall structure of the program. As a
consequence, users have been able to solve problems far outside the
application range originally anticipated by the \dealii{} authors, by
combining building blocks in unexpected and creative ways. On the other
hand, a number of tutorial programs (see
Section~\ref{sec:tool-for-a-large-community}) illustrate how the parts of
\dealii{} \textit{can} be assembled into typical finite element programs.

\subsection{Do not reinvent wheels}
\label{subsec:no_new_wheels} 

Building efficient finite element codes requires tools from a remarkably
broad range of disciplines, ranging across (i) continuous mathematics, such
as in the analysis behind the derivation of error estimators or
approximation results; (ii) discrete and combinatorial mathematics, e.g.,
in the graph algorithms used to partition meshes for parallel computations;
(iii) geometry, for example in the description of meshes with curved
boundaries or on surfaces; (iv) linear algebra for the formulation and
solution of linear systems; (v) computer science concepts related to
parallel computing as well as the design of software as a whole. Other
areas also show up, for example the visualization of data, along with
questions of how to best present data.

No scientific computing project has the manpower and breadth of expertise
to address all of these areas with equal attention to the state of the art.
Thus, a project has to decide to either use only rudimentary algorithms in
some areas, or to use external packages for certain tasks. \dealii{} has
chosen the second route, relying on other software for pre- and
post-processing (i.e., for mesh generation and visualization) and
interfacing with a large number of other software libraries for linear
algebra operations, parallelization, I/O via XML and HDF5, and many other
tasks. The latest \dealii{} version (9.1 at the time of writing) lists 26
other packages with which it interfaces \cite{dealII91}.

This approach has advantages and disadvantages: It allows providing much
more state-of-the-art functionality than we could otherwise. On the other
hand, it requires writing wrappers that may not always expose all options
an underlying package may offer. Furthermore, dealing with large numbers of
dependencies has a substantial cost to both developers and users and is
generally not very well liked. We comment on this in
Section~\ref{subsec:dependency_hell}.

\subsection{Dimension-independent programming}
\label{subsec:dimension_independent_programming}

The way we write partial differential equations today is generally
independent of the dimension we are in. For example, the definition of the
bilinear form for the weak formulation of the Laplace equation is commonly
written as
\begin{equation}
  \label{eq:bilinear_laplace}
  a(u,v) := (\nabla u,\nabla v)_\Omega =
  \int_\Omega \nabla u \cdot \nabla v \, \text{d}x,
\end{equation}
in which the gradient is a $d$-dimensional vector, the dot product
represents a sum over $d$ components, and the integral stretches over a
$d$-dimensional domain $\Omega\subset \R^d$. The actual value of $d$ does
not matter.

It would be nice for many reasons if we could reflect this independence in
implementations of the finite element method. For example, such a scheme
makes it easier to read code because there is a 1:1 relationship with
mathematical notation; it also allows writing code only once, so that it
can be tested using relatively cheap 2d simulations and then used for
production runs in 3d without having to develop and debug a second version.
Earlier finite element libraries, such as DiffPack \cite{BL97}, did so by
equipping essentially every class with a member variable that represents
the dimension. Several modern finite element libraries, such as libMesh
\cite{Kirk2006}, implement a hybrid approach where some classes, such as
\texttt{libMesh::FE}, are templated on the spatial dimension but others,
like \texttt{libMesh::FEMap}, are not. These approaches work but have two
disadvantages: (i) Finite element codes contain an incredible number of
loops over $i=1,\ldots,d$,%
\footnote{For example, \dealii{} has some 3,000 loops that are terminated
in some way by a condition that depends on a constant expression involving
$d$.}
and many of these are in the hot paths of typical execution scenarios;
since the dimension $d$ is a run-time variable in this system, none of
these loops can be unrolled by the compiler. (ii) Memory allocation for
$d$-dimensional vectors (such as the gradients in
\eqref{eq:bilinear_laplace}) must either occur on the heap using
dynamic addressing or use a fixed-size array that is large enough for all
supported spatial dimensions. Other libraries have made the dimension $d$ a
single global constant selected through the build system to avoid these
issues; however, such a system does not allow mixing 2d and 3d simulations,
for example for coupled bulk-surface models.

\dealii{} instead equips many classes with an integer-valued template
argument, in a technique called ``dimension-independent programming''
\cite{Ban00i}. For example, points in $d$-dimensional space are represented
by a class that can be thought of as follows (with many details omitted):
\begin{lstlisting}
  template <int dim> class Point {
    private:   double coordinates[dim];
    public:    double operator[] (const unsigned int i);
  };
\end{lstlisting}
Here, the dimension of the object is known at compile time, allowing the
compiler to unroll and vectorize loops, as well as to allocate the
\texttt{coordinates} array on the stack without wasting space in lower
dimensions. Furthermore, it is possible to use both \texttt{Point<2>} and
\texttt{Point<3>} in the same program.

As most \dealii{} classes have such template arguments, it is possible to
write code that describes things such as the bilinear form of a partial
differential equation in a way that almost exactly resembles mathematical
notation, and compiles to the appropriate code in whatever dimension is
eventually selected. The following code snippet is taken from the tutorial
program
step-4\footnote{\url{https://www.dealii.org/current/doxygen/deal.II/step_4.html}}
and assembles the matrix corresponding to $a(\cdot,\cdot)$ and the vector
corresponding to a right-hand side term
$l(v):=\int_\Omega v\,f\, \mathrm{d}x$
on one cell:
\begin{lstlisting}
for (unsigned int q = 0; q < n_q_points; ++q)
  for (unsigned int i = 0; i < dofs_per_cell; ++i) 
    {
      for (unsigned int j = 0; j < dofs_per_cell; ++j)
        cell_matrix(i, j) +=
          fe_values.shape_grad(i, q) *
          fe_values.shape_grad(j, q) *
          fe_values.JxW(q);
      const Point<dim> x_q = fe_values.quadrature_point(q);
      cell_rhs(i) += fe_values.shape_value(i, q) *
                     right_hand_side.value(x_q) *
                     fe_values.JxW(q);
    }
\end{lstlisting}
The code runs in any dimension and is a literal translation of the
mathematical notation obtained by substituting $\varphi_i(\mathbf x_q)$
with \texttt{fe\_values.shape\_value(i, q)}, $\nabla \varphi_i(\mathbf
x_q)$ with \texttt{fe\_values.shape\_grad(i, q)}, and noting that (with
numerical quadrature) \texttt{fe\_values.JxW(q)} corresponds to the
$\text{d}x$ in the integral. Here, $\mathbf x_q$ denotes the location (in
real space) of the $q$th quadrature point.

\subsection{Iterator-based programming}

\changed{\dealii{} provides access to cells, faces, and vertices via
\textit{iterators}.} Indeed, iterators are flexible because they don't have
to point to an \textit{object} that has member variables (e.g., a cell that
stores the indices or coordinates of its vertices, its material and
subdomain id, etc.); rather, iterators can point to ``accessor'' objects
that store nothing except whatever information is necessary to
\textit{retrieve} pieces of data about a cell or face.

\changed{In particular, accessors} enable the use of far more complicated
data structures than simple arrays of structures. Indeed, the way \dealii{}
stores data is generally in the form of structures of arrays, rather than
arrays of structures, as this leads to substantially better cache locality
\cite{Henning2007}: Loops over all cells rarely access \textit{all} of the
information that is available for each cell, but typically access the same
pieces of data for each cell visited one after the other.

Furthermore, iterators and accessors avoid having to expose the internal
data structures used in \dealii{} classes -- a benefit in maintaining and
optimizing software over the course of many years. As a consequence,
\dealii{} also uses iterator- and accessor-based designs for many other
classes, including sparsity patterns, matrices, index sets, and others.
\changed{This design paradigm enables both the now idiomatic use of
C++11-style range-based for loops, but also the notion of applying a kernel
(often a lambda function) to all elements of a collection.}

\dealii{} supports accessing the subobjects of a cell, such as its faces or
(in 3d) its lines, by returning iterators with different template
parameters:
\begin{lstlisting}
template <int dim, int spacedim>
class CellAccessor : public TriaAccessor<dim, dim, spacedim> {
public:
  TriaIterator<TriaAccessor<dim - 1, dim, spacedim>>
  face(const unsigned int i) const;
};
\end{lstlisting}
The dimensionality of the current structure (\texttt{structdim}), the
topological dimension of the mesh (\texttt{dim}), and the dimension
of the space in which the mesh is embedded (\texttt{spacedim}) are all described by compile-time
constants. One can write \texttt{cell->face(0)} to obtain an iterator for
accessing the  first face of the given cell (so the structure dimension,
the first template argument, is decremented by one). This technique
provides a dimension-independent way of accessing the faces of an element
without the need to construct a proxy element or, in fact, storing any
information in a face-based data structure.

\subsection{Large-scale parallelism}

With the demise of the exponential increase of computing speed of
individual processor cores in the early 2000s, it has become clear that the
numerical solution of complex, three-dimensional PDEs will only ever be
possible by using parallel computing. Hence, \dealii{}  supports serial
computations for prototyping as well as parallelization on both
workstations and clusters, and provides an upgrade path between the two.

Many operations inside \dealii{} are parallelized using task-based
programming (currently via the Threading Building Blocks library
\cite{Reinders07}), with higher level abstractions building on these
concepts \cite{TKB16}. This approach already makes efficient use of
shared-memory systems for many common operations. Beyond this, \dealii{}
also provides distributed memory parallelization of essentially all
operations, using MPI \cite{MPI15} and libraries built on top of MPI
\cite{BangerthBursteddeHeisterKronbichler11, trilinos, trilinos-web-page,
petsc-web-page, petsc-user-ref, p4est, Burstedde2018}. This has allowed for
the creation of programs that make efficient use of machines with hundreds
of thousands of processor cores, see Section~\ref{sec:applications}.

\subsection{Interoperability of all features}
\label{sec:interop}

As outlined above, we see \dealii{} as a library with flexible and
re-usable building blocks. From a practical perspective, these ought to all
work together: If a user wants to switch from a sequential to a parallel
mesh, \changed{they expect} that the finite element class used before will
continue to be usable.

In practice, providing this kind of interoperability leads to a
combinatorial matrix of features that need to be implemented, tested, and
documented if, for example, different triangulation classes had different
requirements of finite element classes. Other examples include supporting
both CPU and GPU computations on both $h$- and $hp$-refined meshes, and
providing every finite element class with the necessary interfaces for
$hp$-adaptivity. Despite these difficulties, one of our design goals is to
allow all combinations of features. We place great emphasis on early
design, code review, and finding the right abstractions in the development
of new features, as these steps make interoperability substantially easier
in the long run.

At the same time, \dealii{} does have combinations of features that do not
(currently) work together. In most of these cases, poor planning can be
attributed to it in retrospect. We will comment on this in
Section~\ref{sec:lessons}.

\subsection{Design for extensibility}

The object-oriented design of \dealii{} enables large features to be added
to the library without making invasive changes to existing classes or
functions. This permits users to replace fundamental parts of the library
with their own implementations.

All finite element classes are ultimately derived from an abstract base
class that specifies the public interface required by the rest of the
library. This is more general than some other libraries such as
\texttt{libMesh}, where \texttt{libMesh::FEFamily} is an enumeration
provided by the library and cannot be changed by the user. Indeed, users
have contributed new and sophisticated elements, such as a new N\'ed\'elec
element that supports arbitrary approximation orders \cite{KRL16}.
Implementations of mappings and geometry descriptions are done in a similar
manner: users can implement these by inheriting
from the \texttt{Mapping} or \texttt{Manifold} classes, respectively.

In other cases, extensibility is provided by template-based generic
programming. In addition to its own linear algebra data structures and
solvers, \dealii{} has wrappers for the linear algebra components of PETSc
\cite{petsc-user-ref}, \trilinos{}'s Epetra and Tpetra subpackages
\cite{trilinos}, cuSPARSE \cite{cusparse}, and Ginkgo
\cite{ginkgo-web-page}. All of these classes are assumed to conform to a
standard interface that permits the use of, e.g., any vector type in the
library with any function that takes a vector argument. For example,
functions that take a finite element coefficient vector, such as
\texttt{VectorTools::integrate\_difference()}, leave the vector type as a
template argument and expect each vector class to implement a member
function \texttt{extract\_subvector\_to()}.

Finally, classes often have nontrivial data dependencies or
interdependencies. For example, \texttt{GridTools::Cache} stores
computationally intensive information about a triangulation. A
\texttt{GridTools::Cache} object will register itself with its associated
\texttt{Triangulation} via a signal/slot mechanism: That is, if the
triangulation is changed, it will inform the cache object, which will then
invalidate relevant information. This pattern is commonly used in \dealii{}
and permits users to express new data dependencies without changing the
implementation of classes in the library.

\subsection{A tool for a large community}
\label{sec:tool-for-a-large-community}
\dealii{} started in 1997 as a tool for one, and shortly after that for
three user-developers, but now serves as the basis for the work of
hundreds, maybe thousands of scientists, producing more than 200
publications per year \cite{dealii-pubs}, in almost any area of science and
engineering one can think of (see also Section~\ref{sec:applications}). It
also has far more developers: 30-50 people have contributed in each of the
most recent releases, and generate 5-10 pull requests per day.
\changed{To ensure quality, we require every change to pass various
continuous integration steps and to undergo rigorous peer review by at
least one of the principal developers.}

The sizes of these communities imply very different requirements than those
that were applied in the early years. For example, we place great emphasis
on compatibility between releases. Likewise, we have built a test suite
with more than \changed{12,000} tests that is run many times a day.
Development versions are almost universally as stable as releases, and the
number of bugs reported on mailing lists and forums is quite small for a
project of this size.

A large user community has many other, often more important, consequences
for a project. In particular, we have long lost the ability to answer
everyone's questions if even a small subset of our user community does not
understand certain concepts or features: If every user had only five
questions per year, we would have a dozen or more questions each day,
consuming resources no volunteer open source project can provide. Rather,
we have placed great emphasis on documentation that guides users through
the process of learning such a tool. This includes the obvious function and
class documentation processed by \texttt{doxygen} \cite{doxygen}. But, it
is also important to explain higher level concepts, and so \dealii{} also
uses \texttt{doxygen} ``modules'' discussing related groups of classes, as
well as a ``tutorial'' of currently more than 60 programs
\cite{dealii-tutorial} that show how the different parts of the library can
be combined in typical finite element codes. The tutorial is also a
\textit{teaching tool} that illustrates many numerical techniques: Each
tutorial program consists of an extensive introduction that discusses the
theoretical background and motivation for the methods used, along with a
thoroughly documented implementation. In the same spirit, we have also
recorded more than 40 hours of video
lectures\footnote{\url{https://www.math.colostate.edu/~bangerth/videos.html}}
(see also Section~\ref{sec:education}) that provide a complementary
perspective as well as interactive demonstrations.

Finally, the ``code gallery'' \cite{dealii-code-gallery} provides a
repository of codes contributed by the user community. They are typically
not as well documented as tutorial programs, but nevertheless can serve as
starting points for others' research. A curated list of publications based
on \dealii{} \cite{dealii-pubs} serves a similar purpose: To showcase what
kinds of applications can be solved using the library.

\subsection{A way to build a community itself}
\label{sec:building-a-community}

Indeed, our approaches to managing a user community can also be seen as an
attempt at \textit{building} a community of learners, users, and
developers. Computational Science and Engineering (CS\,\&\,E) is not an
established discipline with a broad base of degree programs, books, tools,
and methods that newcomers to the field can rely on -- rather, it is a
dynamic and new field \cite{Rde2018} in which many are recent entries and
most are self-taught. Providing concrete, well-documented use cases for
others to learn from, as well as nuclei for learning communities (for
example through forums where people can ask questions of their own and find
answers to others') are important tools to broaden the knowledge base of
CS\,\&\,E practitioners.


\section{Features}
\label{sec:features}

\begin{figure}
    \centering
    \scalefont{0.75}
    \begin{tikzpicture}[auto, node distance=2cm,>=latex']
      \node [core, name=tria] (tria) {\class{Triangulation}};
      \node [core, right of=tria, node distance=3.2cm] (fe) {\class{FiniteElement}};
      \node [core, right of=fe, node distance=2.5cm] (mapping) {\class{Mapping}};
      \node [core, right of=mapping, node distance=2.3cm] (quadrature) {\class{Quadrature}};
      \node [core, below=2em of fe] (dh) {\class{DoFHandler}};
      \node [core, below=2em of mapping] (fevalues) {\class{FEValues}};
      \node [core, below right=2em and -2em of dh] (systems) {Linear systems};
      \node [core, below=1cm of systems] (solvers) {Linear solvers};
      \node [core, below=1cm of solvers] (output) {Graphical output};
      \node [core, above right=2em and 0.5cm of tria] (manifold) {\class{Manifold}};

      \draw[->,line-important] (tria) -- (dh);
      \draw[->,line-important] (fe) -- (dh);
      \draw[->,line-important] (fe) -- (fevalues);
      \draw[->,line-important] (mapping) -- (fevalues);
      \draw[->,line-important] (quadrature) -- (fevalues);
      \draw[->,line-important] (dh) -- (systems);
      \draw[->,line-important] (fevalues) -- (systems);
      \draw[->,line-important] (systems) -- (solvers);
      \draw[->,line-important] (solvers) -- (output);
      \draw[->,line-important] (manifold) -- (tria);
      \draw[->,line-important] (manifold) -- (mapping);

      \node [library, right=3cm of systems] (petsc) {\library{PETSc}};
      \node [library, below=1em of petsc] (trilinos)  {\library{Trilinos}};
      \node [library, below=1em of trilinos] (cuda)  {\library{CUDA}};
      \node [library, below=1em of cuda] (linalgmisc)  {\library{...}};

      \node [library, below left=1em and 3em of systems] (umfpack)  {\library{UMFPACK}};
      \node [library, below=1em of umfpack] (linalg)  {\library{...}};

      \node [library, right=3cm of manifold] (opencascade) {\library{OpenCASCADE}};

      \draw[->,line-optional] (petsc) -- (systems);
      \draw[->,line-optional] (petsc) -- (solvers);

      \draw[->,line-optional] (trilinos) -- (systems);
      \draw[->,line-optional] (trilinos) -- (solvers);

      \draw[->,line-optional] (cuda) -- (systems);
      \draw[->,line-optional] (cuda) -- (solvers);

      \draw[->,line-optional] (linalgmisc) -- (systems);
      \draw[->,line-optional] (linalgmisc) -- (solvers);

      \draw[->,line-optional] (umfpack) -- (solvers);
      \draw[->,line-optional] (linalg) -- (solvers);

      \draw[->,line-optional] (opencascade) -- (manifold);

      \node[application, below=1.5cm of tria] (gmsh) {\application{gmsh}};
      \node[application, below left=0.5cm and 0.0cm of output] (visit)  {\application{VisIt}};
      \node[application, right=1em of visit] (paraview) {\application{ParaView}};
      \node[application, right=1em of paraview] (viz) {\application{...}};

      \draw[<->,line-optional] (gmsh) -- (tria);
      \draw[->,line-optional] (output) -- (visit);
      \draw[->,line-optional] (output) -- (paraview);
      \draw[->,line-optional] (output) -- (viz);
    \end{tikzpicture}
    \caption{Core components of \dealii{} and interplay with some external
      libraries.}
    \label{fig:core-elements}
\end{figure}
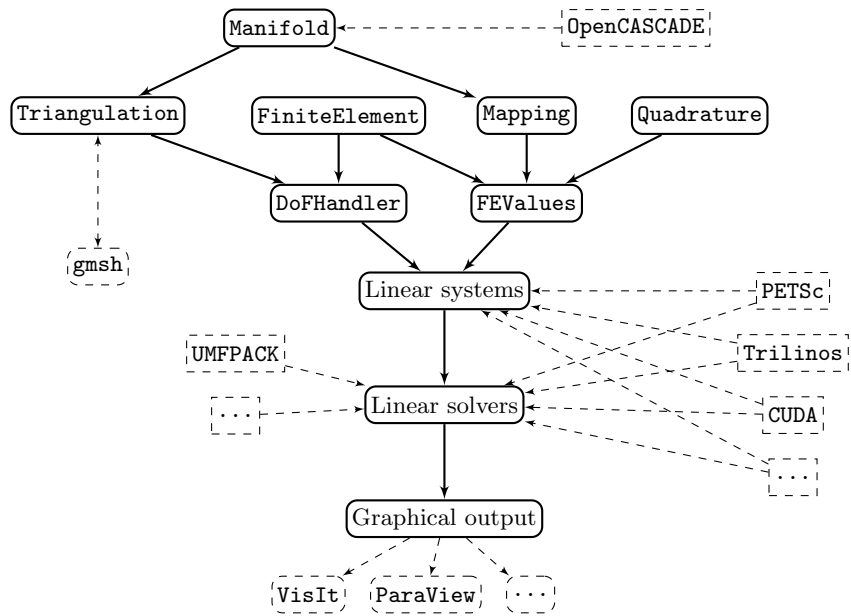

Having discussed what we want to achieve with \dealii{}, let us now turn to
a discussion of the features the library offers.
Fig.~\ref{fig:core-elements} provides an overview of the biggest building
blocks of \dealii{} and their interplay. Each box references a
\textit{concept} that is, in most cases, implemented in several different
ways -- either as classes derived from a common base class (e.g., in the
case of the finite elements, mappings, and quadrature classes), as
independent classes using a generic interface (as is the case for the
\texttt{DoFHandler} and linear algebra concepts), or a combination
thereof. The figure also references a few of the external libraries
\dealii{} can interface with.

Rather than discussing each of the components of this graph in detail (this
is done in the technical documentation of \dealii{}), we focus instead on
several overarching themes and considerations.

\subsection{Sequential, shared, and distributed triangulations}
\label{subsec:triangulations}

A key concept in all finite element codes is the mesh, i.e., a collection
of cells that cover the domain in question. For historical reasons, meshes
are often called ``triangulations'', even if -- as is the case in \dealii{}
-- they consist of line segments (in 1d), quadrilaterals (in 2d), or
hexahedra (in 3d).

\dealii{}'s restriction to quadrilaterals and hexahedra was originally
motivated by the observation that, for equal numbers of degrees of freedom,
finite element solutions with tensor product elements tend to be more
accurate than those using the same approximation order spaces on triangles
or tetrahedra (the additional accuracy comes from the extra cross terms,
e.g., the $x y$ basis function in a 2d $Q_1$ element).

This limitation simplifies a large number of algorithms: For example, many
common elements and quadrature formulas can be implemented in a
dimension-independent way with arbitrary approximation order by exploiting
a tensor-product structure. Similarly, (isotropic) \(h\)-adaptive mesh
refinement is executed in essentially the same way in 2d and 3d by
splitting parent cells into \(2^\mathrm{dim}\) child cells. This kind of
adaptive mesh refinement in turn provides a convenient setting for
implementing geometric multigrid algorithms: the construction of
prolongation and restriction operators (as well as level construction) is
greatly simplified in the case of hierarchical mesh refinement. Finally, a
large number of topological quantities become compile-time constants: The
number of faces, edges, and vertices of cells are all fixed and known if
all cells with spatial dimension \texttt{dim} are the same shape.
\changed{At the same time, it is true that it is often difficult to
generate good quadrilateral or hexahedral meshes (other than the trivial
subdivision of a triangle or tetrahedron into 3 or 4 quadrilaterals or
hexahedra, using a tool such as \library{tethex}). This clearly limits the
applicability of \dealii{} in some applications, though our experience is
that this really only significantly affects the use of complex, 3d
geometries. In many other, practical situations, relatively coarse meshes
can either be generated by hand or by mesh generators and then refined
automatically to fit the geometry of the object using the techniques
described in Section~\ref{sec:geometry_abstractions}.}

\dealii{} currently has three triangulation classes: sequential, (parallel)
shared, and (parallel) distributed. The latter two partition the mesh among
MPI processes, making the parallel solution of partial differential
equations possible. The difference between the shared and the distributed
triangulation is what each process stores: In the shared case, each process
stores the entire triangulation -- wasteful in terms of memory and only
scalable to around 100 processes, but useful when dealing with problems
that require knowledge of the entire mesh on each process (as in boundary
element methods). In contrast, the distributed triangulation stores the
coarse mesh everywhere, which is then refined hierarchically, and each
process only stores the subset of locally owned cells of this refined mesh,
along with ghost cells surrounding the locally owned cells.

The parallel distributed mesh implementation in \dealii{} is
algorithmically much more involved than the parallel shared mesh
\changed{\cite{BangerthBursteddeHeisterKronbichler11}}. On the other hand,
it provides a distributed data structure that has been shown to scale to
very large numbers of MPI processes and unknowns: We have demonstrated
computations on up to 304,128 processes and up to $2\cdot
10^{12}$~\changed{\cite{Arndt19}} unknowns, far beyond what is necessary to
solve most problems in practice \changed{today}.

For all three of the triangulations mentioned above, the dimensionality of
the mesh may differ from the dimensionality of the space in which it lives.
This allows for the solution of equations on surfaces embedded in higher-dimensional spaces. Examples where this is useful are the use of boundary
element methods, but also modeling surface processes on solids and fluids
(possibly coupled to models of the enclosed bulk medium) such as surface
tension, erosion, or diffusion on membranes.

\subsection{Geometry abstractions}
\label{sec:geometry_abstractions}

\begin{floatingfigure}[r]{0.40\textwidth}
    \centering
    \includegraphics[width=0.37\textwidth]{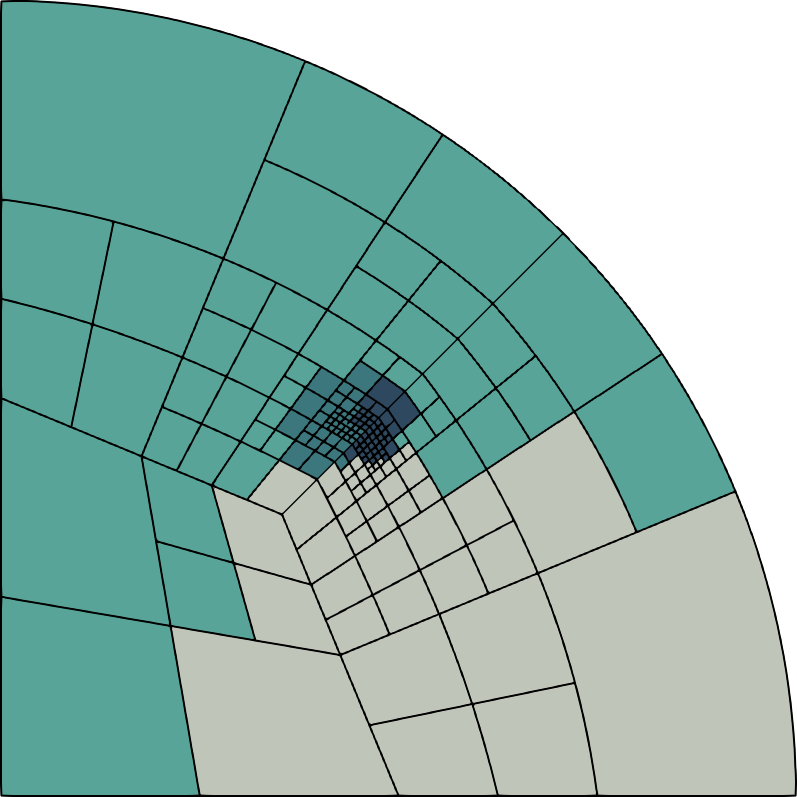}
    \caption{A mesh adaptively refined from 3 coarse cells with hanging
      nodes and curved faces. The geometry description represents the
      boundary as an exact circle. This geometry is extended into the
      interior using transfinite interpolation. Cells are colored based on
      a partitioning onto four processes.}
    \label{fig:fancy-mesh}
\end{floatingfigure}

\vspace{-1em}
A key feature of \dealii{} is support for adaptive mesh refinement and
coarsening. Refining cells in highly distorted or curved domains is a
challenging problem that requires both a description of the underlying
geometry and algorithms for using this information to create child cells
that are no more distorted than their parent cell. It is also often
necessary to propagate information from a curved boundary into the interior
of a triangulation to achieve a well-conditioned discretization. These
problems are especially important when computing solutions on surfaces
embedded in higher dimensional surfaces since, in this context, all cells
will usually be curved. Accurate geometry descriptions are also critical
for higher-order discretizations because they enable one to use a boundary
description whose order of accuracy matches that of the finite element
space (e.g., isoparametric and isogeometric finite elements).

The implementation of this functionality in \dealii{} is based on the
language of differential geometry and resides in classes inheriting from
\texttt{Manifold}. These manifold descriptions are used in (i) placing
new vertices upon mesh refinement, (ii) computing
normal and tangential vectors to the boundary, and (iii) defining the
mapping from the reference cell to a concrete cell of the mesh; they also
appear in a number of other operations.  Isoparametric finite elements use
\texttt{Manifold} objects to compute the coordinates of support points that
lie along curved faces as well as support points on the interior of the
cell (through the \texttt{MappingQ} and\texttt{MappingFEField} classes).
Similarly,  the abstraction provided by manifolds enables the use of
isogeometric mappings (through the \texttt{MappingManifold} class), where
the exact geometry is used directly to define the mapping from the
reference cell to a concrete cell of the mesh.

When a volumetric description of the geometry is not available, CAD models
can be used to represent the geometry of the boundaries, and transfinite
interpolation \cite{Gordon82} may be used to extend surface geometry
descriptions into the interior of the domain. A complete discussion of
these issues is provided in \cite{geometry_paper}; an example is shown in
Fig.~\ref{fig:fancy-mesh}.

\vspace{1em} 

\subsection{\texorpdfstring{$h$}{h} and \texorpdfstring{$hp$}{hp} adaptivity}

\dealii{} was originally designed as a library supporting adaptive mesh
refinement, i.e., $h$-adaptivity, using hanging nodes as a means to deal
with differing cells sizes. It was later extended to also support
$hp$-adaptivity whereby one can also individually select the \changed{finite element,
and therefore the polynomial
degree,} on each cell \cite{BangerthKayserHerold2007}. While technically
difficult to implement -- in particular for continuous elements, in 3d, and
in parallel -- supporting $hp$-adaptivity has also enabled a host of other,
unexpected applications if one allows for different cells to use entirely
different kinds of finite elements. In particular, the introduction of the
\texttt{FE\_Nothing} element that represents a space with no degrees of
freedom (i.e., only consisting of the zero function), has allowed
representing multiphysics applications in which some equations are only
posed on parts of the domain.

\subsection{Finite elements, mappings, and evaluating them}
\label{sec:variety_of_fe}

Over the past few decades, an entire zoo of finite element spaces has been
described in the literature.
\changed{The publication by Arnold and Logg in \cite{arnold2014periodic}
and references therein summarizes the most commonly used ones, but there
are many others (a subset illustrating the breadth of elements can be found
in \cite{Adini1961,Argyris1968,CloughTocher,DSSY99,Arnold2006}).} \dealii{}
implements a substantial fraction of \changed{these}, starting with the
common continuous and discontinuous Lagrange elements, but also the
N\'ed\'elec, Raviart-Thomas, Arnold-Boffi-Falk, Brezzi-Douglas-Marini,
Bernardi-Raugel, Bernstein, P1-nonconforming, Rannacher-Turek, and other
spaces. There are also implementations of more specialized cases: Finite
element spaces enriched by bubble functions or using additional weights (as
necessary for the XFEM approach); monomial bases; unmapped non-parametric
bases; and finite element spaces defined only on the faces of cells. Almost
all of these are available with arbitrary polynomial degree.

Most of these spaces are defined on the reference cell $[0,1]^d$ and need
to be mapped to each cell of the triangulation. This is done using one of
several implementations of the \texttt{Mapping} base class: The usual
polynomial mappings based on Lagrange interpolation points; a specialized
mapping that can be used in case all cells are rectangles/boxes with axes
parallel to a Cartesian coordinate system; and mappings that are ``exact''
in the sense that they respect the underlying manifold description of the
domain's geometry.

An important realization, explored in more detail in \cite[Section
3.3]{BangerthHartmannKanschat2007}, is that while finite element spaces and
mappings describe \textit{functions}, in practice they only need to be
\textit{evaluated at individual points} because the integrals used in the
finite element method are approximated by quadrature. The \texttt{FEValues}
class provides an interface that allows querying shape functions,
transformed to the current cell, at a set of evaluation points. Furthermore, because
the quadrature points (in reference coordinates) are typically the same for
a loop over all cells, \texttt{FEValues} pre-computes and caches as much
information as possible, substantially accelerating the computation on a
sequence of cells in a transparent manner. Similar classes exist for
evaluating shape functions and derivatives on faces.

For problems with multiple solution components -- say, a flow problem whose
solution consists of $d$ vector components for a velocity plus one for the
pressure -- it is often convenient to either consider the combined finite
element space of all components (e.g., to correctly size the linear
system); or to only consider its restriction to a scalar component, the $d$
components of a vector, or the $d^2$ (or $d(d+1)/2$) components of a
(symmetric) tensor (e.g., when assembling the bilinear form). In \dealii{},
the latter is facilitated by ``extractor'' objects that, when applied to an
\texttt{FEValues} object, yield a ``view'' of the selected finite element
space.

\subsection{Geometric multigrid}

The ability to solve large linear systems \changed{stemming from finite
element discretizations} requires solvers that are both optimal in
complexity (i.e., $O(N)$ where $N$ is the number of unknowns) and scale
well in parallel. Only multigrid methods, either algebraic or geometric,
are known to fulfill this requirement \changed{for elliptic problems.}

\dealii{} has long supported and successfully used algebraic multigrid
(AMG) methods as a preconditioner through Trilinos' ML/MueLu packages and
PETSc's hypre interfaces; for examples, see
\changed{\cite{BangerthBursteddeHeisterKronbichler11,KHB12}}. However, AMG
setup costs become prohibitive for very large problems and core
counts~\cite{clevenger_stokes}. Consequently, \dealii{} also supports
geometric multigrid methods; originally written for sequential computations
on adaptive meshes in 1999, it now also supports parallelization via
multithreading and MPI \cite{clevenger_par_gmg}.

While algebraic multigrid methods can often be treated as black-box
preconditioners, the implementation of the geometric multigrid algorithm
consists of a flexible framework with various options and customization
points: First, it integrates seamlessly with existing linear algebra
classes (PETSc, \trilinos{}) or matrix-free smoothers and transfer
operators (see Section~\ref{subsec:matrixfree}). Second, there is a large
collection of smoothers that can be used, from matrix-based operations
(e.g., Jacobi or SSOR, including various parallel variants), to Schwarz
smoothers that allow local smoothers on cells or patches of cells.
Similarly, coarse solvers can be provided through various means, including
switching to algebraic multigrid.

This framework has been shown to scale to extremely large problems, see
Section~\ref{sec:applications}. Customization points and block composition
of solvers allows using the multigrid framework for a wide variety of PDEs.

\subsection{Matrix-free operators}\label{subsec:matrixfree}
It is well-understood that matrix-based linear solvers can no longer
adequately use the computational power of modern CPUs or GPUs because of
the disparity between the high cost of transferring data from memory and the
speed of floating point operations. Consequently, for computational efficiency,
linear solvers need to find a better balance between precomputing and
storing data, versus computing more information on the fly.

To this end, \dealii{} contains functionality for matrix-free computations
that merges the assembly and solver steps. In this paradigm, a global
sparse matrix is never built and linear systems are only solved by the
action of the underlying linear operator on a vector via the integrals in
the weak form, such as $(\nabla v, \nabla u)_{\Omega}$ from
Equation~\eqref{eq:bilinear_laplace}. Since the information needed to
compute the integrals in terms of the geometry and possibly some
coefficients is much smaller in memory than a sparse matrix that encodes
the coupling of every unknown to the others, a matrix-free approach has the
potential to avoid the limitations of slow memory access. It also often
allows one to fit bigger problems into memory. Matrix-free setups offer
more optimization possibilities than sparse matrices because one can choose
what should be pre-computed and stored; the fundamental question being
whether any increase in arithmetic operations costs less than what one can
win from the reduced memory transfer. On today's hardware, it turns out to
be particular useful if one can exploit the tensor-product form of shape
functions \cite{KronbichlerKormann2012,KronbichlerKormann2019}. In
\dealii{}, we provide matrix-free capabilities on a subset of element
types; currently, these are in particular the tensor-product continuous and
discontinuous elements.

\changed{To map these arithmetically intensive operations efficiently onto
modern hardware,} \dealii{} supports single-instruction/multiple data
(SIMD) instructions, such as SSE, AVX, and AVX-512 on x86 hardware or
AltiVec on IBM's Power. \changed{Given that the compilers'
auto-vectorization is typically only applicable to operations in the
innermost loops, which are often operations over the components of a tensor
in $d$ spatial dimension, loop reorganization towards
array-of-struct-of-array data layouts promise much better
performance~\cite{KronbichlerKormann2019}. In \dealii{} we provide such
layouts by a single class called \texttt{VectorizedArray} that internally
applies intrinsics for the various instruction set extensions. To the user
code,} the usual operator overloading infrastructure makes it behave
similarly to the built-in types \texttt{double} or \texttt{float},
\changed{while abstracting away the detail of the innermost array
dimension. All use cases can transparently invoke the most beneficial SIMD
array width for the hardware, without compromising code portability among
CPU architectures.}

\subsection{Support for graphics processing units}

Similarly, \dealii{} also supports operations on Graphics Processing Units
(GPUs), due to their superior floating point capabilities and memory
bandwidth for a given power \changed{consumption}. Multiple programming
models are available for running code on a GPU, e.g., CUDA~\cite{cuda},
Kokkos~\cite{Carter2014}, RAJA~\cite{raja,Hornung2015},
OCCA~\cite{medina2014}, and OpenMP~\cite{openmp}. \dealii{} builds on CUDA,
due to the maturity of the language; while this limits the use of these
features to Nvidia GPUs, the extension to AMD GPUs using \emph{hipify}
\cite{rocm} is underway. GPU capabilities in \dealii{} consist of
matrix-based classes wrapping functionality of the cuSPARSE and cuSOLVER
libraries \cite{cusparse,cusolver}, and matrix-free support.

For matrix-based computations, the linear system needs to be assembled on
the host first. The data is then moved to the device, converted to a format
usable with cuSPARSE, and finally handed to the preconditioners and direct
solvers from cuSOLVER. \dealii{}'s own iterative solvers can also be used,
with all linear algebra operations performed on the GPU.

In contrast, \dealii{}'s CUDA matrix-free framework does not require
computation of the operator on the host first. Instead, \changed{finite
element integrals} are evaluated directly on the device. This implies that
users will need to write part of their code in CUDA. Given most users' lack
of CUDA experience, we have minimized the amount of necessary CUDA code by
only requiring a user interface with a single functor that can be
implemented using code close to what the user would write when doing
matrix-free implementations on the host. While we have striven to have the
interface for GPU matrix-free framework to be as close as possible to the
CPU matrix-free framework, they are not identical. Aside from the obvious
difference of having the user writing a functor annotated with
\texttt{\_\_device\_\_}, the more fundamental difference is due to the way
the parallelization is done.  On the CPU, each thread works on a separate
chunk of cells while on the GPU, each thread works on a different
degree-of-freedom. This is due to the fact that the GPU has much more
parallelism available but not as much memory per thread. Details of the
implementation are discussed in \cite{Ljungkvist2017}.

The matrix-free framework can also take advantage of multiple GPUs through
the use of MPI via \dealii{}'s own data structures and solvers. Data can be
transferred between different GPUs on different systems either via the
host's memory or, if an implementation supports it, without this detour.

\subsection{Assisted differentiation and linearization}

Many realistic applications use formulations derived from complicated
energy functionals or residuals. Examples are large deformation
formulations of materials with nonlinear constitutive laws, fluid-structure
interaction problems, or chemically reacting flows. In such cases, deriving
and consistently implementing the bilinear forms and right hand sides that
characterize each nonlinear step is often fraught with opportunities for
error. Additionally, validating an implementation with no analytical
solution or verifying that the convergence rate of a nonlinear solution
scheme is optimal also becomes extremely challenging.

In many such cases, workflows can be substantially simplified by
integrating tools for automatic differentiation (AD). To this end,
\dealii{} leverages ADOL-C \cite{Griewank1996a,Walther2009a} and Sacado
\cite{Bartlett2006a,Gay2012a,Phipps2012a} to automatically compute first
and second derivatives. On the finite element level, this approach allows
computing the Newton matrix from a residual (``linearization''), or both
the residual and its linearization can be determined from an energy
functional. The same can analogously be done on the quadrature point level.
This framework abstracts away the specialized function calls and operations
that each of the supported automatic differentiation (AD) libraries (and
the underlying number types) requires and offers a clear and unified
interface to the users.

Complementing the AD framework is an interface to the high performance
Computer Algebra System (CAS) and symbolic toolbox SymEngine
\cite{Certik2013a} that performs symbolic calculations on scalar types
using common C++ syntax through operator overloading. Valid operations not
only include standard mathematical operations and symbolic differentiation,
but also comparison, logical, and conditional operations. Having integrated
the scalar SymEngine wrapper class, called \texttt{Expression}, into the
pre-existing \texttt{Tensor} and \texttt{SymmetricTensor} classes, we have
also equipped the framework to perform tensorial computations and,
subsequently, symbolic tensor differentiation as tensors are commonly used
in the definition of residuals and energy functionals. Its flexibility
makes the symbolic framework well suited to perform specialized tasks that
have a complex code path and require either partial or total derivatives to
be computed.

The framework above lays the foundation to include other exciting features
in the future. In particular, opportunities include offloading symbolic
computations to a just-in-time compiler for more rapid execution of
computed operations, and symbolic finite element level assembly and
linearization in a similar spirit to that previously described for the AD
framework.


\section{Lessons learned from the development of complex software}
\label{sec:lessons}

Having discussed design goals and available functionality in the previous two
sections, it may be interesting to also put all of this development into
perspective: What have we learned about the development of a complex
scientific software library intended for a large user community?

In \cite{BH13}, we have previously given some answers on what we think
makes scientific software libraries successful. Let us here summarize some
of the points made there \changed{and explain how they relate to \dealii{} specifically,}
but also -- and in particular -- discuss a few of
the things we know are difficult or for which we do not know how to do them
well.

\subsection{A success: Testing}

As mentioned in Section \ref{sec:tool-for-a-large-community}, having an
extensive testsuite is essential to providing stable functionality. This is
in particular true in view of the continuous growth in the number of
possibilities to configure the library, using different compilers,
different dependencies, and hardware platforms. To this end, \dealii{}'s
test suite with \changed{more than 12,000} tests is run continuously with a wide cross-section
of all configuration combinations. Compiling and running a substantial
fraction of the test suite is part of the continuous integration hook for
each patch, and success is required before a patch can be merged.

\subsection{A challenge: Dependency hell and installation support}
\label{subsec:dependency_hell}

A consequence of the design decision to not implement functionality for
which specialized external libraries exist (see
Section~\ref{subsec:no_new_wheels}), is that \dealii{} relies on
\textit{many} other packages. Almost all of these (and their corresponding
wrappers) are optional, but most advanced \dealii{}-based projects likely still depend on
some. Thus, users are required to install these external projects by hand,
rely on pre-installed versions made available by their system
administrators, or use tools that facilitate building scientific software
such as Spack \cite{Spack}.

Configuration management is notoriously complicated. Many scientific
packages, to this day, use installation procedures based on Makefile
snippets or autoconf, neither of which export the details of their
installation for downstream packages. Many also have broken installations
in which, for example, shared libraries do not record which other shared
libraries they depend upon, leaving it to downstream projects to figure out
what to link with. Furthermore, every system seems to be different: Not
just between Linux, \changed{macOS,} and Windows, but even within each of these
operating systems, there are substantial differences in what is available,
where, and how. Finally, the projects \dealii{} can interface with make
incompatible changes between versions, either requiring supporting multiple
versions at once, or requiring users to use a specific version.

All other scientific software projects we know of struggle with this
``dependency hell'' and ``version hell''; no particular good and widely
usable solutions appear to exist. We try to address these issues by
providing installation scripts that automatically download optional
dependencies, such as \texttt{candi} \cite{candi}, and by working through
package managers (of package repositories such as Debian or Ubuntu, or
  source based installers such as Spack \cite{Spack} and the xSDK
\cite{xsdk} environment that builds on it).

\subsection{A success: Backward compatibility issues}

Just like other packages, \dealii{} carries legacy functionality for which
we have found better solutions over time. This is particularly relevant
because libraries such as \dealii{} expose such large interfaces to users:
hundreds of classes and thousands of functions are accessible to users.
Replacing or changing any of them would break downstream codes, and prevent
upgrade paths for our users from one version to the next.

We have always tried to minimize incompatible changes that would impact
users. Where this cannot be avoided, we deprecate functionality in one
version and remove it in the next. Using compiler features, the use of
deprecated functionality is possible, but triggers a compiler warning
alerting users to the need to eventually update their code. This appears to
be working: We very rarely hear complaints about lack of backward
compatibility.

\subsection{A challenge: Support for changing architectures}

Computer architectures are changing and, with the advent of GPUs and Xeon
Phis, also more diverse. Some architectures have multiple levels of
heterogeneity: Nvidia's Volta GPU architecture is itself accelerated by
``tensor'' cores. Any open source project will struggle with natively
supporting all of them. \dealii{} does support a limited number of
architectures (x86, POWER, and Nvidia GPUs) natively. For other
architectures, \dealii{} will need to rely on third-party libraries such as
the Tpetra package of \trilinos{}~\cite{trilinos} that uses Kokkos
\cite{Carter2014} for performance portability.

\subsection{A challenge: Supporting ``different'' discretizations}

Not all discretizations considered widely useful are as simple as standard
isoparametric Lagrange finite elements mapped to each element via a
polynomial mapping. Isogeometric analysis (IGA), finite elements based on
Catmull-Clark's subdivision surfaces, the extended finite element method
(XFEM), and certain types of enriched finite elements all do not fit into
this scheme. In those cases, the degrees of freedom cannot be thought of as
being associated with a specific mesh object, but rather a collection of
such objects (for example, a patch of cells). For these, shape functions
may not be defined based on some ``reference cell'', they may extend beyond
just one cell and its immediate neighbors, and the number of shape
functions supported on a given cell may depend on the topology of the mesh
around that element (as is the case for Catmull-Clark's finite
elements).

Some of these schemes are difficult to press into the current design of the
\texttt{FiniteElement} class, and of the \texttt{FEValues} class that
provides point values and derivatives (see Section
\ref{sec:variety_of_fe}). Similarly, enumeration of degrees of freedom
(DoFs) poses challenges. To address these issues, the description of the
finite element space, represented by the \texttt{FiniteElement} class, will
have to learn to provide information on how many of those ``non-local''
DoFs there are on a given mesh. This information will then need to be used
by the \texttt{DoFHandler} class responsible for globally enumerating DoFs.
Additional complications arise in the MPI-parallel context where ownership
of a non-local DoF can no longer be determined based on which cells each
process owns because these degrees of freedom are no longer associated with
a particular cell. One might also need to have a ghost layer thicker than a
single cell.

No library can support everything, but there is an ongoing effort to add
basic support for such ``non-local'' DoFs to the \dealii{} library.

\subsection{A challenge and a success: Interoperability of ``single-use'' features}

Some of \dealii{}'s features have turned out to not to be interoperable
with other parts of the library. They were typically written for a single
use case, and often only implemented or understood by a single person.
Their \changed{focus on specific topics} hinders adoption by a larger
number of users; those who do \changed{use them} find themselves
frustrated by lack of support for these features in other parts of the
library. These features are also often poorly documented, and are
overrepresented among questions on the online forums.

In hindsight, these features were contributed with good intentions, but
became a maintenance problem especially if the contributor later walked
away from the project. We have learned from this: The modules in question
predate the time when every patch had to pass peer review, and large
patches are now often extensively discussed for design choices and
interoperability before they are accepted. Our standards for documentation
are also far higher today than they were before every patch was reviewed.

\subsection{A success \changed{and a challenge}: Organizing large volunteer projects}

A scientific project with hundreds or thousands of users, and dozens of
contributors to each release, can be considered a success. \changed{We also
know} of some 1,500 publications from essentially all areas in the sciences
and engineering that use \dealii{} \cite{dealii-pubs}; only a small
fraction of these was authored by the principal developers of the project.

\changed{Getting to these numbers required a lot of deliberate work and
thinking about how best to use our human resources: A substantial fraction
of our effort is spent on supporting users via online forums,%
\footnote{\changed{An overview of resources for help is available at
\url{https://www.dealii.org/participate.html}, including links to the help
forum, issue tracker, and the project's GitHub repository.}}
but importantly also on providing extensive documentation and other
supports to avoid having to answer everyone's questions on the public
forums. It also includes taking the perspective of users into consideration
during the development process.} Finally, we consciously focus on growing
users into developers, by encouraging them to contribute and providing them
with mentoring on their patches.

\changed{At the same time, there are also challenges when dealing with
actual humans. These include dealing with ``difficult'' community members,
not burning any one person in the project out by overloading them with
answering forum questions, giving newer project members space to find a
niche, and not expecting new contributors to write perfect patches with
extensive documentation or tests before a pull request can be merged. An
important distinction from other projects appears to be that among the
\dealii{} principal developers, none seem to be particularly territorial
about their code and all are quite happy to let others reshape the code
they wrote. This enables newcomers to contribute without having to fear
that they step on old-timers' feet, and has allowed us to avoid the
tendency of projects to balkanize into individual developers' territories
that only they are allowed to touch.}


\section{Education}
\label{sec:education}

As already mentioned in Sections~\ref{sec:tool-for-a-large-community} and
\ref{sec:building-a-community}, libraries such as \dealii{} do not exist in
a vacuum. Rather, they are surrounded by user communities requiring
education: Help and documentation resources at varying levels, and basic
training in the underlying numerical methods and software development. But,
a project such as \dealii{} also \textit{enables} educational
opportunities: It is a tool to reach communities who may be interested in
computational science. Finally, it is also a tool to research \textit{how
best to teach CS\,\&\,E}.

Indeed, many of us have leveraged \dealii{} for educational purposes: We
know that it is used in teaching finite element courses at many
universities around the world, and several of us have taught short courses
based on it on many continents. Even more broadly used is a collection of
currently 67 video lectures that one of us has recorded at KAMU, a
professional television studio. These videos -- hosted on
YouTube%
\footnote{\url{https://www.youtube.com/playlist?list=PLdy04DoEepEwRGMbxwmPTmNBD5jFvhlZM}}
as well as the Chinese bilibili video hosting platform%
\footnote{\url{https://www.bilibili.com/video/av57103047/}}
-- have collectively received more than 140,000 views, indicating robust
user-community interest in learning about the computational science
concepts discussed, as well as in the interactive demonstrations showing how to
build, use, and develop software based on \dealii{}.

The original purpose of the video lectures was to facilitate flipped
classroom teaching in which students learn the material before class,
allowing the instructor more time for interaction with students. We have
found that this approach works well, both anecdotally from the perspective
of those who have used this approach in their own teaching, but also backed
up by rigorous educational research \cite{ZB14,ZBB19}.


\section{Applications}
\label{sec:applications}

\dealii{} is used by hundreds or thousands of researchers in essentially
every field of the sciences and engineering, as shown by the large number
of publications that build on it \cite{dealii-pubs} -- too many to even try
and summarize. Most of these publications use codes written for a specific
purpose \changed{but not publicly available}. However, the project
website at \url{https://www.dealii.org} also links to a number of large
projects built on \dealii{} in the geosciences, radiation transport, the material sciences,
fuel cell modeling, wave propagation, and multiphysics modeling, that
have themselves grown substantial user
communities -- in some cases with hundreds of users of their own.
\changed{Finally, we are aware of on the order of 10-20 industry projects
that build on \dealii{}, and believe that this is likely a substantial
underestimate. Several of the principal developers have industry projects
themselves, ranging from the simulation of ship hulls, wave propagation in
the aerospace field, resonances of industrial membranes, to 3d printing.}

Many of the papers referenced \changed{in the list of papers that use
\dealii{}} \cite{dealii-pubs} provide excellent examples of the breadth and
depth of applicability of \dealii{}. Furthermore, the publications
mentioned at the end of the Introduction also discuss in great detail
individual features. Rather than duplicate this information,
\changed{let us here only summarize} one example of large-scale
computations \changed{in the following sub-section.}

\subsection{\changed{Large-scale computations}}

\pgfplotstableread{
nodes  8m        66m       524m      4b        34b      268b      2.1t
1      3.173e-02 2.643e-01 nan       nan       nan       nan       nan
2      1.657e-02 1.378e-01 1.076e+00 nan       nan       nan       nan
4      9.169e-03 6.950e-02 5.427e-01 nan       nan       nan       nan
8      5.486e-03 3.461e-02 2.758e-01 nan       nan       nan       nan
16     3.705e-03 1.754e-02 1.412e-01 1.094e+00 nan       nan       nan
32     3.060e-03 1.013e-02 7.158e-02 5.517e-01 nan       nan       nan
64     2.732e-03 6.747e-03 3.651e-02 2.827e-01 nan       nan       nan
128    2.543e-03 4.623e-03 1.869e-02 1.442e-01 1.103e+00 nan       nan
256    3.308e-03 5.148e-03 1.180e-02 7.782e-02 5.749e-01 nan       nan
396    3.473e-03 4.211e-03 9.919e-03 5.489e-02 4.038e-01 nan       nan
792    3.490e-03 3.973e-03 7.171e-03 3.009e-02 2.058e-01 1.521e+00 nan
1584   3.538e-03 4.298e-03 6.290e-03 1.698e-02 1.051e-01 7.746e-01 nan
3168   3.993e-03 5.108e-03 5.904e-03 1.091e-02 5.812e-02 3.918e-01 nan
6336   nan       nan       nan       7.865e-03 3.064e-02 1.988e-01 1.506e+00
}\tableMGDG

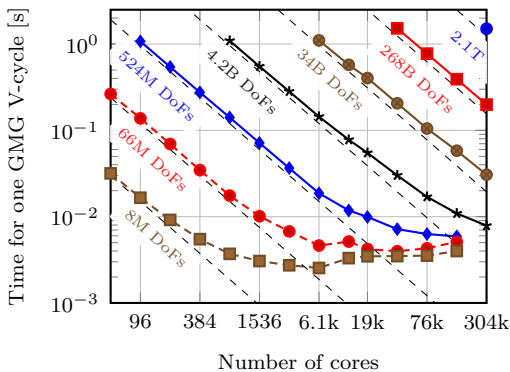
\begin{floatingfigure}{0.5\textwidth}
\centering
  \begin{tikzpicture}
    \begin{loglogaxis}[
        width=0.48\textwidth,
        height=0.4\textwidth,
        xlabel={Number of cores},
        ylabel={Time for one GMG V-cycle [s]},
        xtick={96,384,1536,6144,19008,76032,304128},
        xticklabels={96,384,1536,6.1k,19k,76k,304k},
        tick label style={font=\scriptsize},
        label style={font=\scriptsize},
        legend style={font=\scriptsize},
        ymin=1e-3, ymax=2.5,
        xmin=48, xmax=304128,
        grid, thick
      ]
      \addplot table[x expr={\thisrowno{0}*48}, y index={7}] {\tableMGDG}  node [pos=0,anchor=west,below=1.5mm,fill=white,opacity=0.9,rotate=-42]{\tiny 2.1T\phantom{ab}};
      \addplot table[x expr={\thisrowno{0}*48}, y index={6}] {\tableMGDG}  node [pos=0.35,anchor=west,below=1.5mm,fill=white,opacity=0.9,rotate=-42]{\tiny 268B DoFs};
      \addplot table[x expr={\thisrowno{0}*48}, y index={5}] {\tableMGDG}  node [pos=0.14,anchor=west,below=1mm,fill=white,opacity=0.9,rotate=-42]{\tiny 34B DoFs};
      \addplot table[x expr={\thisrowno{0}*48}, y index={4}] {\tableMGDG}  node [pos=0.1,anchor=west,below=1mm,fill=white,opacity=0.9,rotate=-42]{\tiny 4.2B DoFs};
      \addplot table[x expr={\thisrowno{0}*48}, y index={3}] {\tableMGDG}  node [pos=0.1,anchor=west,below=1mm,fill=white,opacity=0.9,rotate=-42]{\tiny 524M DoFs};
      \addplot table[x expr={\thisrowno{0}*48}, y index={2}] {\tableMGDG}  node [pos=0.18,anchor=west,below=1mm,fill=white,opacity=0.9,rotate=-42]{\tiny 66M DoFs};
      \addplot table[x expr={\thisrowno{0}*48}, y index={1}] {\tableMGDG}  node [pos=0.20,anchor=west,below=1mm,fill=white,opacity=0.9,rotate=-42]{\tiny 8M DoFs};
      \addplot[dashed,black,thin] coordinates {
        (48,0.03)
        (152064,0.03/3168)
      };
      \addplot[dashed,black,thin] coordinates {
        (48,0.03*8)
        (152064,0.03/3168*8)
      };
      \addplot[dashed,black,thin] coordinates {
        (48,0.03*64)
        (152064,0.03/3168*64)
      };
      \addplot[dashed,black,thin] coordinates {
        (48,0.03*512)
        (152064,0.03/3168*512)
     };
      \addplot[dashed,black,thin] coordinates {
        (48,0.03*4096)
        (304128,0.03/6336*4096)
      };
      \addplot[dashed,black,thin] coordinates {
        (48,0.03*32768)
        (304128,0.03/6336*32768)
      };
      \addplot[dashed,black,thin] coordinates {
        (48,0.03*262144)
        (304128,0.03/6336*262144)
      };
    \end{loglogaxis}
  \end{tikzpicture}
  \vspace{-2em}
  \caption{\it Parallel scaling of a GMG V-cycle on up to 304,128 cores and
    up to \num{2e12} unknowns. Dashed lines show ideal scaling.}
  \label{fig:scaling_mg}
\end{floatingfigure}

The computations shown in this section are based on
\cite{Arndt19,KronbichlerWall2018}. Figure~\ref{fig:scaling_mg} shows a
scaling experiment of a geometric multigrid V-cycle on the SuperMUC-NG
machine in Garching, Germany, with up to 304,128 Intel Xeon Platinum 8174
cores. The equation being solved here is the three-dimensional Laplace
equation; however, solving this or related equations also serves as the
basis for block preconditioners for more complicated equations such as
the Stokes and Navier-Stokes equations, as well as for elastic and plastic
models, and the performance shown here is therefore indicative also of
solvers for more complex problems. The example shown uses a symmetric
interior penalty discontinuous Galerkin discretization with polynomial
degree $p=4$ on a hyper-rectangle. The multigrid hierarchy
involves a switch from the discontinuous space to the associated continuous
finite element space with $p=4$ on the same mesh \cite{Antonietti2017} and
then progresses to coarser mesh levels until a coarse solver is invoked on
a $2\times 1\times 1$ mesh. Chebyshev smoothing of degree $6$ for pre- and
post-smoothing is used on all levels. For operator evaluation, the
matrix-free infrastructure (Section~\ref{subsec:matrixfree}) with AVX-512
vectorization is used. The multigrid V-cycle is run in single precision to
increase throughput, which is the typical usage setting when combined with
some double-precision correction~\cite{KronbichlerLjungkvist19}.

This setup achieves a multigrid convergence rate of about $0.03$, i.e.,
\changed{the residual is reduced} by 3 orders of magnitude with just two
V-cycle applications.  The largest problem with \num{2.15e12} unknowns runs
with \changed{(single-precision)} arithmetic throughput of \changed{5.9}
PFlop/s and a memory throughput of 1.2 PB/s (187 GB/s per node), and is
primarily limited by the memory bandwidth. The combination of these
features then leads to a solver for the Laplace equation that can solve a
problem with trillions of unknowns in just a few seconds -- opening the
door for solving much more complex problems that use the Laplace equation
as one block in the preconditioner.


\section{\changed{Vision and future directions}}
\label{sec:future}

\changed{Like many volunteer projects, \dealii{} does not have a formal
roadmap that drives development. Rather, most new features stem from
individuals' needs that arise in their research.
That said, we have the
following aspirational goals:
\begin{itemize}
    \item
      To provide state-of-the-art tools for a broad variety of PDEs.
    \item
      To have excellent documentation to enable anyone to use the library.
    \item
      To enable fast and scalable algorithms from laptops to the largest
      supercomputers, including new architectures.
    \item
      To create an open, inclusive, participatory community providing users
      and developers with the resources they need.
\end{itemize}
In support of these goals, there are a number of technical areas in which concrete work is currently ongoing and that we will discuss briefly in the following.
\paragraph*{Efficient computations on the largest machines} As the results of the previous section show, \dealii{}-based codes have been used on the full scale of some of the biggest machines available today, with more than $10^{12}$ degrees of freedom. Another example, among several others, is the code discussed in detail in \cite{Motamarri2020}; in both of these cases, computations are run on more than 100,000 cores. Yet, experience shows that every time one goes to larger numbers of cores or more degrees of freedom, one encounters new bottlenecks. We continue to address these by carefully timing parts of our code base, but also by addressing algorithms that we know can not scale indefinitely. An example is the implementation and optimization of matrix-free methods, and their current incorporation into multigrid methods that scale to very large problems \cite{clevenger_par_gmg}.
\paragraph*{GPU support} Similarly, we know that for the foreseeable future, the largest compute clusters will draw the majority of their compute power from GPUs and similar devices. Yet, supporting this very different programming model is difficult, and needs to be hidden from user code wherever possible. \dealii{} has supports for GPUs, and strives to make it available more broadly.
\paragraph*{Parallel $hp$ adaptivity} $hp$ refinement has been part of \dealii{} for the past 15 years, and parallel distributed meshes for more than 10. Yet, the combination of these two features has only been available for the last year or so. We continue to work on making the necessary connections, and in particular devising practical strategies for ensuring that the resulting computations are well load-balanced.
\paragraph*{Tutorials and other pieces of the documentation}
Providing users with adequate documentation is a continual challenge. There are never enough tutorial programs to cover all of the applications our users have in mind, nor do the existing ones cover the changing programming models we use in view of large-scale parallelism, GPUs, or evolving discretization methods. Likewise, despite 20 years of work, function and class documentation never seems to be adequate to answer all user questions. As a consequence, a substantial fraction of commits over the years has always been to improve the documentation, and there are 2-4 new tutorial programs every year. This trend will surely continue indefinitely.
}


\section{Conclusions}
\label{sec:conclusions}

In this contribution, we have outlined the design criteria and
functionality of the \dealii{} finite element library. Like many other
scientific software projects, \dealii{} started as a small project within
one lab, with no intention of reaching beyond that point; however, it has
now grown into a successful and world-wide project that is used in hundreds
or thousands of research projects, with nearly a dozen principal developers
who spend a substantial fraction of their time on the continued development
of the package. This change from a project for a few user-developers to a
community project brings with it not only an explosion in functionality (as
discussed in Section~\ref{sec:features}), but also a reckoning on how such
software can be developed: It requires an agreement on the design
principles that guide continuing development (see Section~\ref{sec:design})
but also a focus on the technical and social challenges a project of this
size brings with it (Section~\ref{sec:lessons}). At the same time, it also
opens up opportunities as a widely used teaching tool
(Section~\ref{sec:education}).

As this article, \changed{and especially Section~\ref{sec:applications}}, made clear, \dealii{} is
no longer a hobbyists' project, but a professionally managed enterprise
whose continued development has, over the years, been supported repeatedly
by a multitude of funding agencies and that has not only built more than a
million lines of C++, but also a vibrant and active user and developer
community.

\subsection*{Acknowledgments.}

W.~Bangerth was partially supported by the National Science Foundation
(NSF) under Award OAC-1835673 as part of the Cyberinfrastructure for
Sustained Scientific Innovation (CSSI) program; by Award DMS-1821210; and
by the Computational Infrastructure in Geodynamics initiative (CIG),
through the National Science Foundation under Awards No.~EAR-0949446 and
EAR-1550901 and The University of California -- Davis.

T.~Heister was partially supported by NSF Award DMS-1901529, OAC-2015848,
EAR-1925575, by the Computational Infrastructure in Geodynamics initiative
(CIG), through the NSF under Award EAR-0949446 and EAR-1550901 and The
University of California -- Davis, and by Technical Data Analysis, Inc.
through US Navy STTR N16A-T003.

L.~Heltai was partially supported by the Italian Ministry of
Instruction, University and Research (MIUR), under the 2017 PRIN
project NA-FROM-PDEs MIUR PE1, 
``Numerical Analysis for Full and Reduced Order Methods for the
efficient and accurate solution of complex systems governed by Partial
Differential Equations''.

M.~Kronbichler was partially supported by the German Research Foundation
(DFG) under the project ``High-order discontinuous Galerkin for the
exa-scale'' (\mbox{ExaDG}) within the priority program ``Software for
Exascale Computing'' (SPPEXA).

M.~Maier was partially supported by ARO MURI Award No.~W911NF-14-1-0247, as
well as NSF Award DMS-1912847.

D.~Wells was partially supported by NSF Award OAC-1450327 as part of the
Cyberinfrastructure for Sustained Scientific Innovation (CSSI) program.

Research sponsored by the Laboratory Directed Research and Development
Program of Oak Ridge National Laboratory, managed by UT-Battelle, LLC, for
the U.S. Department of Energy.

The Gauss Centre for Supercomputing e.V.~(\texttt{www.gauss-centre.eu}) is
acknowledged for providing computing time on the GCS Supercomputer
SuperMUC-NG at Leibniz Supercomputing Centre through project id pr83te.

Clemson University is acknowledged for generous allotment of compute time
on Palmetto cluster.

All authors gratefully acknowledge the feedback and support by the vast
user community of \dealii{}, as well as the many contributions made by
people from around the world. This project could not survive without its
user and developer communities!


\bibliographystyle{elsarticle}

{\footnotesize\bibliography{paper.bib}}

\end{document}